\documentclass[review,preprint,12pt]{elsarticle}

\usepackage{amssymb,amsmath,bm,comment}

\journal{}

\begin{document}

\begin{frontmatter}



\title{Synchronized molecular-dynamics simulation for the thermal lubrication of a polymeric liquid between parallel plates}


\author{Shugo YASUDA\corref{cor1}}
\address{Graduate School of Simulation Studies, University of Hyogo}
\cortext[cor1]{Email: yasuda@sim.u-hyogo.ac.jp, Tel/Fax +81(0)783031990}
\author{Ryoichi YAMAMOTO}
\address{Department of Chemical Engineering, Kyoto University}

\begin{abstract}
The Synchronized Molecular-Dynamics simulation which was recently proposed by authors [Phys. Rev. X {\bf 4}, 041011 (2014)] 
is applied to the analysis of polymer lubrication between parallel plates.
The rheological properties, conformational change of polymer chains, and temperature rise due to the viscous heating are investigated
with changing the values of thermal conductivity of the polymeric liquid.
It is found that at a small applied shear stress on the plate, 
the temperature of polymeric liquid only slightly increases in inverse proportion to the thermal conductivity 
and the apparent viscosity of polymeric liquid is not much affected by changing the thermal conductivity.
However, at a large shear stress, the transitional behaviors of the polymeric liquid occur due to the interplay of the shear deformation and viscous heating by changing the thermal conductivity.
This transition is characterized by the Nahme-Griffith number $Na$ which is defined as the ratio of the viscous heating to the thermal conduction 
at a characteristic temperature.
When the Nahme-Griffith number exceeds the unity,
the temperature of polymeric liquid increases rapidly and the apparent viscosity also exponentially decreases as the thermal conductivity decreases.
The conformation of polymer chains is stretched and aligned by the shear flow for $Na<1$, but the coherent structure becomes disturbed by the thermal motion of molecules for $Na>1$.
\end{abstract}


\end{frontmatter}


\section{Introduction}
\label{sec1}
Molecular dynamics (MD) simulations are used to investigate the materials properties and internal structures of the complex liquids and soft matters.
In MD simulations, the dynamics of all the model molecules which compose of the materials are calculated.
The material properties and internal structures of the materials are analyzed from the results of the MD simulations.
The MD simulation is applicable for a wide variety of complex liquids and soft matters consist of the complicated molecules
such as the polymer molecules and the biological molecules.
However, due to the enormous computational time to calculate the dynamics of all the molecules, the MD simulation is not yet 
applicable to problems which concern mesoscopic and macroscopic scales motions far beyond the molecular size, where 
the dynamics of molecules are strongly coupled to the macroscopic transports of the complex liquids.
At microscopic scales, the macroscopic transports are usually ignored and the MD simulation is performed under a certain ideal environment.
To develop a simulation technology to analyze the molecular dynamics coupled with the macroscopic transports for the complex liquids 
is a significant challenge from both a scientific and engineering point of view.
Multiscale modeling is a promising candidate to address this challenge.

The multiscale simulation for the flow behaviors of complex fluids was first advanced in the CONNFFESSIT approach for polymeric liquids by Laso and
 \"Ottinger\cite{art:93LO,art:95FLO,art:97LPO}, where the local stress in the fluid solver is calculated using the microscopic simulation instead of 
using any constitutive relations. 
The GENERIC approach is also presented for the non-isothermal polymeric flows by making important corrections and clarifications to the CONNFFESSIT scheme.\cite{art:99DEO} 
The strategy exploited in the CONNFFESSIT approach is also introduced into the heterogeneous multiscale modeling (HMM), which was proposed by E and Enquist,\cite{art:03EE} 
as a general methodology for the efficient numerical computation of problems with multiscale characteristics. 
HMM has been applied to various problems, such as the simple polymeric flow\cite{art:05RE}, coarsening of a binary polymer blend\cite{art:11MD} and the channel flow of a 
simple Lennard-Jones liquid\cite{art:13BLR}. 
The equation-free multiscale computation proposed by Kevrekids et al. is also based on a similar idea and has been applied to various problems.\cite{art:03KGHKRT, art:09KS} 
De et al. proposed the scale-bridging method, which can correctly reproduce the memory effect of a polymeric liquid, and demonstrated the non-linear viscoelastic behavior of 
a polymeric liquid in slab and cylindrical geometries.\cite{art:06DFSKK, art:13D} 
The multiscale simulation for polymeric flows with the advection of memory in two and three dimensions was developed by Murashima and Taniguchi.\cite{MT2010,MT2011,MT2012} 
We have also developed a multiscale simulation coupling of MD and transport equations. 
The multiscale method was first developed for simple fluids\cite{art:08YY} and subsequently extended to polymeric liquids with the memory effect.\cite{art:09YY,art:10YY,art:11YY,art:13MYTY}
Recently, we proposed the synchronized molecular dynamics (SMD) method in which the multiscale method was extended to treat the coupled heat and momentum transfer of complex liquids.\cite{art:14YY}

In the previous study, the SMD method was applied to the polymer lubrication generating the viscous heating between parallel plates.
The rheological properties, conformation of polymer chains, and temperature rise due to the viscous heating were investigated for various applied shear stresses on the plate.
We found that an interesting transitional behavior of the conformation of polymer chains occurs with a rapid temperature rise for a large applied shear stress
due to the coupling of the shear deformation and heat generation under the shear flow.

In this study, we investigate the effects of changing the thermal conductivity of the polymeric liquid on the behaviors of rheological properties, 
conformation of polymer chains, and heat generation in the polymer lubrication between parallel plates.
In the following, we first describe the problem considered in the present paper. 
The simulation method is briefly explained after the presentation of the problem. 
The SMD simulation of polymer lubrication is performed, and the results are discussed; 
these results are mainly the rheological properties and the coupling of the conformations of the polymer chains with heat generation under shear flows. 
Finally, a short summary is given.

\begin{figure*}[tbp]
\begin{center}
\includegraphics[scale=0.9]{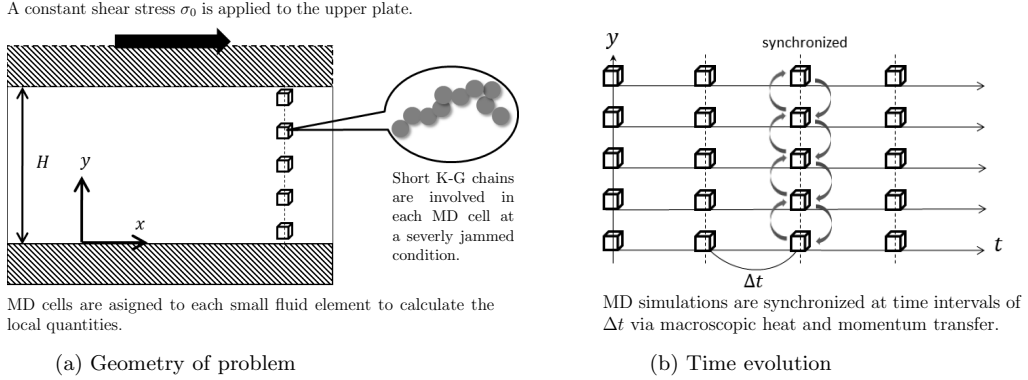}
\caption{%
The schematic of the geometry of the problem (a) and the time evolution (b).
This figure is reprinted from Ref. \cite{art:14YY}.
}\label{fig_problem}
\end{center}
\end{figure*}
\section{Problem and Method}
We consider a polymeric liquid contained in a gap of width $H$ between parallel plates with a constant temperature $T_0$ (see Fig. \ref{fig_problem}(a)). 
The polymeric liquid is composed of short Kremer-Grest chains\cite{art:90KG} of ten beads, in which all of the bead particles interact with a truncated Lennard-Jones potential defined by
\begin{equation}
U_{\rm LJ}(r)=\left\{
\begin{array}{l r}
4\epsilon\left[
\left(\frac{\sigma}{r}\right)^{12}
-\left(\frac{\sigma}{r}\right)^{6}
\right]
+\epsilon,
&\quad (r\le 2^{1/6}\sigma),\\
0,
&\quad (r > 2^{1/6}\sigma),
\end{array}
\right.
\end{equation}
and consecutive beads on each chain are connected by an anharmonic spring potential,
\begin{equation}
U_{\rm F}(r)=-\frac{1}{2}k_c R_0^2 \ln
\left[
1-\left(\frac{r}{R_0}\right)^2
\right],
\end{equation}
with $k_c$=30$\epsilon/\sigma^2$ and $R_0$=$1.5\sigma$. The polymeric liquid is in a quiescent state with a uniform density $\rho_0$ and a uniform temperature $T_0$ before a time $t=0$. Hereafter, the $y$-axis is perpendicular to the parallel plates, and the boundaries between the upper and lower plates and the polymeric liquid are located at $y=H$ and 0, respectively. 
The upper plate starts to move in the $x$ direction with a constant shear stress $\sigma_0$ at a time $t$=0, while the lower plate is at rest.

The macroscopic behavior of the polymeric liquid is described by the following transport equations:
\begin{subequations}\label{eq_macro}
\begin{align}
\rho_0\frac{\partial v_x}{\partial t}&=\frac{\partial \sigma_{xy}}{\partial y},
\label{eq_flow}\\
\rho_0\frac{\partial e}{\partial t}&=\sigma_{xy}\dot\gamma - \lambda \frac{\partial^2 T}{\partial y^2},
\label{eq_ene}
\end{align}
\end{subequations}
where $v_\alpha$ is the velocity, $\sigma_{\alpha\beta}$ is the stress tensor, $e$ is the internal energy per unit mass, and $\dot \gamma$ is the shear rate, i.e., $\dot \gamma=\partial v_x/\partial y$. Hereafter, the subscripts $\alpha$, $\beta$, and $\gamma$ represent the index in Cartesian coordinates, i.e., \{$\alpha$,$\beta$,$\gamma$\}$\in$\{$x$,$y$,$z$\}. Here, we assume that the macroscopic quantities are uniform in the $x$ and $z$ directions, $\partial /\partial x$=$\partial /\partial z$=0, and the density of the polymeric liquid is constant.
Fourier's law for a heat flux with a constant and uniform thermal conductivity $\lambda$ is also considered in Eq. (\ref{eq_ene}). Note that the thermal conductivity of polymeric liquids is anisotropic under shear flows in general\cite{art:90BB,art:96OP,art:96BC,art:97BCB}, and some experimental studies have reported that the linear stress-thermal relation between the stress tensor and thermal conductivity holds.\cite{art:01VSIGB,art:04SVBBS,art:12SVG,art:13GSV}
However, in the present study, we only consider the isotropic thermal conductivity as the first step because the effect of shear thinning of the viscosity is thought to be more crucial to viscous heating under strong shear flows than that of the anisotropy of the thermal conductivity.
We also assume that the velocity and temperature of the polymeric liquid are the same as those of the plates at the boundaries, i.e., the non-slip and non-temperature-jump boundary conditions.

The effect of viscous heating is estimated using the ratio of the first and second terms in Eq. (\ref{eq_ene}) to be $\sigma_0\dot\Gamma H^2/\lambda \Delta T_0$. Here, $\dot\Gamma$ is the gross shear rate of the system, which is defined by the ratio of the velocity of the upper plate $v_w$ to the width of the gap $H$, $\dot \Gamma=v_w/H$, and $\Delta T_0$ is a characteristic temperature rise for the polymeric liquid. In the present problem, we consider a characteristic temperature necessary to substantially change the viscosity of the polymeric liquid, i.e., $\Delta T_0=|\eta_0/(\partial \eta_0/\partial T_0)|$, where $\eta_0$ is the characteristic viscosity of the polymeric liquid at a temperature of $T_0$. Thus, the Nahme-Griffith number $Na$, defined as 
\begin{equation}\label{eq_nahme-griffith}
Na=\frac{\sigma_0\dot\Gamma H^2}{\lambda|\partial \log(\eta_0)/\partial T_0|^{-1}} 
\end{equation}
represents the effect of viscous heating on the changes in the rheological properties.\cite{book:87BAH,art:08PMM} Usually, 
in lubrication systems and in high-speed processing operations with polymeric liquids, 
the Nahme-Griffith number is not negligibly small because of the large viscosity and the small thermal conductivity of the polymeric liquid.\cite{art:08PMM} 
For example, when a lubrication oil in a gap with a width of 10 $\mu$m is subjected to shear deformation with a strain rate of $1\times 10^4$ $\rm s^{-1}$, 
the Nahme-Griffith number is estimated to be $Na\gtrsim 0.1$. 
Thus, the rheological properties of the lubricant in such micro devices must be significantly affected not only by the large velocity gradient but also by the temperature increase caused by local viscous heating. 
To predict the rheological properties of the polymeric liquid in these systems, one must consider the temperature variation in Eq. (\ref{eq_ene}) coupled with Eq. (\ref{eq_flow}).

The viscous heating is one of the most fundamental and important problems in the lubrication systems.
However, it is difficult to analyze the behaviors of polymer chains under the coupling of strong shear flows and viscous heating with using the conventional simulation technologies.
As is seen in Eq. \ref{eq_nahme-griffith}, the effect of viscous heating on the temperature and behaviors of polymer chains is enhanced as the system size $H$ is large.
Thus, the thermal lubrication is one of the typical problems that the macroscopic transports and microscopic molecular dynamics are strongly interacted each other.
We solve this problem by using the Synchronized Molecular-Dynamics (SMD) simulation which was proposed recently by authors\cite{art:14YY}.

In this paper, we briefly explain the SMD method.
The molecular dynamics cells are assigned to small fluid elements with a interval $\Delta x$ and calculate the local stresses, temperatures, and conformations of polymer chains in each MD cell by using the non-equilibrium molecular dynamics (NEMD) simulations according to the local shear rates $\dot \gamma$. (see Fig. \ref{fig_problem}(b)).
The NEMD simulations are performed independently in each MD cell for a certain time duration $\Delta t$, but at each time interval $\Delta t$, the NEMD simulations are synchronized to satisfy the macroscopic heat- and momentum-transport equations Eq. (\ref{eq_macro}).
Thus, the couplings of the shear flow, viscous heading, and conformational dynamics of polymer chains are autonomously calculated in each MD cell with satisfying the macroscopic heat- and momentum-transfers.

\section{Results}
In this study, we investigate the effect of changing the thermal conductivity $\lambda$ of the polymeric liquid on the rheological property, conformation of polymer chains, and temperature rise.
We carry out the SMD simulations with changing the value of the thermal conductivity $\lambda$.
(In the previous study\cite{art:14YY}, a fixed value of the thermal conductivity is only considered while the applied shear stress to the upper plate $\sigma_0$ is varied variously.)
Hereafter, we measure the length, time, temperature and density in units of $\sigma$, $\tau_0=\sqrt{m\sigma^2/\epsilon}$, $\epsilon/k_B$, and $m/\sigma^3$, respectively. 
Here, $k_B$ is the Boltzmann constant, and $m$ is the mass of the LJ particle. 
In the following simulations, the density of the polymeric liquid is fixed to be $\rho_0=1$, and the temperature of the plates and the width of the gap between the plates are fixed to be $T_0=0.2$ and $H=2500$, respectively.
At this density $\rho_0$ and this temperature $T_0$, the conformation of the bead particles becomes severely jammed and results in complicated rheological properties.\cite{art:11YY,art:02YO}
Two values of the applied shear stress on the upper plate $\sigma_0$ are considered in this study, i.e., $\sigma_0$=0.01 and 0.05.
The SMD simulations are performed with $\lambda=$50, 100, 150, 200, and 300 for $\sigma_0=$0.01 and $\lambda=$50, 75, 100, 125, 150, 200, and 500 for $\sigma_0=0.05$, respectively.
In the following, we present quantities averaged for a long period of time at the steady state, in which the shear stress is spatially uniform and the time derivative of the local
internal energy, i.e., the left-hand side of Eq. (\ref{eq_ene}), is negligible.

\begin{figure}[tbp]
\includegraphics[scale=0.9]{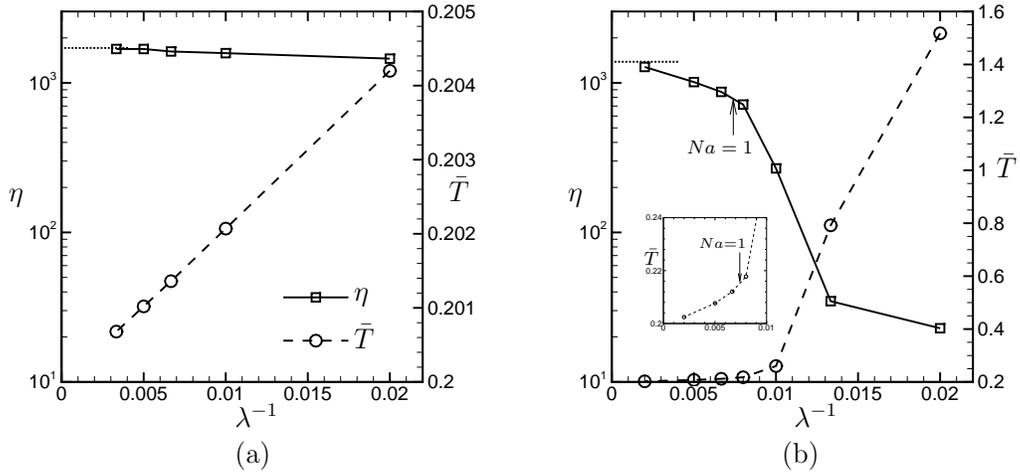}
\caption{
The apparent viscosity $\eta$, defined as $\eta=\sigma_0/\dot\Gamma$, and the spatial average of the temperature $\bar T$ 
as functions of the inverse of thermal conductivity $\lambda^{-1}$ for the applied shear stresses 
(a) $\sigma_0=0.01$ and (b) $\sigma_0=0.01$, respectively.
In (b), inset shows the spatial average of temperature as the function of $\lambda^{-1}$ for large thermal conductivities.
The horizontal dotted lines on the left vertical axises in both figures indicate the viscosities for the infinitely large thermal conductivity, $\lambda^{-1}\rightarrow 0$.
In (b), the vertical arrows show the points at which the Nahme-Griffith number $Na$ equals unity.
}\label{fig_lambda_eta_temp}
\end{figure}
Figure \ref{fig_lambda_eta_temp} shows the dependency of the apparent viscosity $\eta$, defined as $\eta=\sigma_0/\dot\Gamma$, on the thermal conductivity $\lambda$
and the spatial average of temperature $\bar T$.
For a small applied shear stress, i.e., $\sigma_0=0.01$, the apparent viscosity only slightly decreases as the thermal conductivity $\lambda$ decreases (the inverse of thermal conductivity
$\lambda^{-1}$ increases).
The average temperature $\bar T$ increases linearly as the function of the inverse of thermal conductivity $\lambda^{-1}$, but the variation is very small; 
the temperature rise is at most 2.5\% of the wall temperature in the figure.
The facts that the variations of apparent viscosity $\eta$ and average temperature $\bar T$ are small and the average temperature $\bar T$ is a linear function of the inverse of thermal conductivity $\lambda^{-1}$ can be explained by the energy transport equation, Eq. (\ref{eq_ene}), for small Nahme-Griffith numbers; 
when the Nahme-Griffith number is small the viscosity of the polymeric liquid is not much affected due to the viscous heating, so that the local viscosities and shear rates are uniform
between the parallel plates and the temperature rise is inversely proportional to the thermal conductivity.
In Fig. \ref{fig_lambda_eta_temp}(a), the Nahme-Griffith number is at most 0.06.

For a large applied shear stress, i.e., $\sigma_0=0.05$, the behaviors of the apparent viscosity and average temperature are quite different from those for a small applied shear stress.
For $\sigma_0=0.05$, the Nahme-Griffith number of the polymeric liquid can be larger than unity, $Na>1$, at small thermal conductivities.
In Fig. \ref{fig_lambda_eta_temp}(b), the points at which the Nahme-Griffith number equals unity are indicated by the vertical arrows and the Nahme-Griffith number is larger than unity
on the right side of the vertical arrow.
It is seen that the average temperature rapidly increases when the Nahme-Griffith number exceeds the unity, and the apparent viscosity also rapidly decreases as the function of the inverse of thermal conductivity.
When the Nahme-Griffith number exceeds the unity, the viscosity substantially deceases due to the viscous heating, so that the shear velocity increases and the viscous heating is enhanced.
This positive feedback causes the rapid changes of the apparent viscosity and average temperature for small thermal conductivities.  

\begin{figure}[tb]
\includegraphics[scale=0.9]{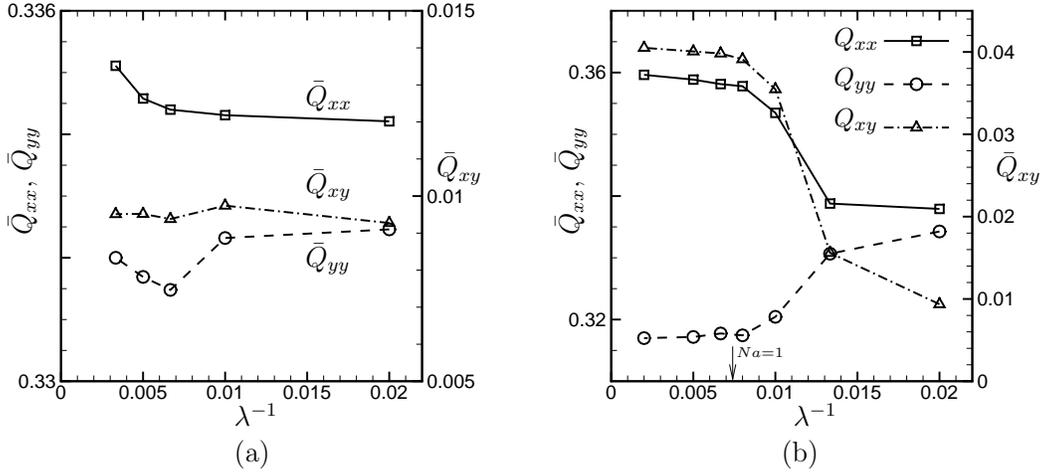}
\caption{
The spatial average of the bond-orientation tensor of the polymer chains $\bar Q_{\alpha\beta}$ as a function of the inverse of thermal conductivity $\lambda^{-1}$
for the applied shear stresses (a) $\sigma_0=0.01$ and (b) $\sigma_0=0.01$, respectively.
In (b), the vertical arrows show the points at which the Nahme-Griffith number $Na$ equals unity.
}\label{fig_lambda_Qxy}
\end{figure}
Figure \ref{fig_lambda_Qxy} shows the conformational changes in polymer chains as a function of the inverse of thermal conductivity.
Here, the bond-orientation tensor $Q_{\alpha\beta}$ is defined as,
\begin{equation}\label{bond_orientation}
Q_{\alpha\beta}=\frac{1}{N_{\rm p}}\sum_{\rm chain}\frac{1}{N_{\rm b}-1}\sum_{j=1}^{N_{\rm b}-1}\frac{b_{j\alpha}}{b_{\rm min}}\frac{b_{j\beta}}{b_{\rm min}},
\end{equation}
where $N_{\rm p}$ is the number of polymer chains in each MD cell, $N_{\rm b}$ is the number of bead particles in a polymer chain, ${\bm b}_j$ for $1\le j\le N_{\rm b}-1$ is the bond vector between consecutive beads in the same chain, and $b_{\rm min}$ is the distance at which the sum $U_{\rm LJ}(r)+U_{\rm F}(r)$ has a minimum and is calculated to be $b_{\rm min}\simeq 0.97$.
As the polymer chains are stretched and aligned by the shear flows driven by the applied shear stress to the upper plate,
the difference of the $xx$- and $yy$-components of the bond orientation tensor, $Q_{xx}-Q_{yy}$, and the $xy$ component of the bond orientation tensor $Q_{xy}$ becomes large.
It is seen that, for a small applied shear stress (Fig. \ref{fig_lambda_Qxy}(a)), the conformations of polymer chains are slightly stretched and aligned by the shear flows.
The variations of each component of the bond orientation to the thermal conductivity are very small; the relative variations are at most 0.3\% for $\bar Q_{xx,\,yy}$ and 5\% for $\bar Q_{xy}$.

For a large applied shear stress (Fig. \ref{fig_lambda_Qxy}(b)), the transitional behaviors occurs around the point at which the Nahme-Griffith number equals unity.
For large thermal conductivities (on the left side of downward arrow), the conformations of polymer chains are highly stretched and aligned 
by the shear flows.
However, for small thermal conductivities (on the right side of downward arrow), the alignment of the polymer chains is disturbed 
due to the thermal motion of molecules 
and the conformation of the polymer chains becomes more isotropic although the shear velocities are enhanced as the temperatures increase.
This indicates that the transition between the regime dominated by the shear flow and dominated by the thermal motion of molecules occurs by changing the thermal conductivity. 

\section{Summary}
We carry out the SMD simulations for the thermal lubrication of a model polymeric liquid composed of short chains between parallel plates.
The SMD simulations for the same problem were also carried out in Ref. \cite{art:14YY}, where the behaviors of the polymeric liquid for various applied shear stresses are investigated.
In the previous study we found that the transitional behaviors occur in the dynamics of polymer chains by changing the applied shear stress;
when the Nahme-Griffith number is small, $Na<1$, the conformation of polymer chains is stretched and aligned by the shear flow, but when the Nahme-Griffith number exceeds unity, $Na>1$, 
the coherent structure becomes disturbed by the thermal motion of molecules.

In the present study, the effects of changing the thermal conductivity of the polymeric liquid on the rheological properties, conformation of polymer chains, and temperature rise due to the viscous heating are investigated.
It is found that at a small applied shear stress on the plate, i.e., $\sigma_0=0.01$, the temperature of polymeric liquid only slightly increases in inverse proportion to the thermal conductivity and the apparent viscosity of polymeric liquid is not much affected by changing the thermal conductivity since the Nahme-Griffith number is very small.
However, at a large shear stress, i.e., $\sigma_0=0.05$, the Nahme-Griffith number of the polymeric liquid can be larger than unity, $Na>1$, 
when the thermal conductivity is sufficiently small.
When the Nahme-Griffith number exceeds unity, the temperature of polymeric liquid increases rapidly and the apparent viscosity exponentially decreases as the thermal conductivity decreases.
The conformation of polymer chains is stretched and aligned by the shear flow for $Na<1$, but the coherent structure becomes disturbed by the thermal motion of molecules for $Na>1$.
Thus, in this study we demonstrate that the transitional behaviors which were found in the previous study also occur by changing the thermal conductivity under a fixed applied shear stress.

\section*{acknowledgement}
This study was financially supported by JSPS KAKENHI Grant Numbers 26790080 and 26247069.
The computations have been performed using the supercomputer of ACCMS, Kyoto University.




\begin{thebibliography}{00}

\bibitem{book:89AT}
M. P. Allen and D. J. Tildesley,
{\it Computer Simulation of Liquids},
(Oxford University Press, New York, 1989).
%
\bibitem{book:08EM}
D. J. Evans and G. Morris,
{\it Statistical mechanics of nonequilibrium liquids},
(Cambridge university press, New York, 2008).
%
\bibitem{book:87BAH}
R. B. Bird, R. C. Armstrong, and O. Hassager,
{\it Dynamics of polymeric liquids} Vol. 1 (John Wiley and Sons, New York, 1987).
%
%
\bibitem{art:93LO}
M. Laso and H. C. \"Ottinger,
``Calculation of viscoelastic flow using molecular models: the CONNFFESSIT approach'',
J. Non-Newtonian Fluid Mech. {\bf 47}, 1 (1993).
%
\bibitem{art:95FLO}
K. Feigl, M. Laso, and H. C. \"Ottinger,
``CONNFFESSIT approach for solving a two-dimensional viscoelastic fluid problem'',
Macromolecules {\bf 28}, 3261 (1995).
%
\bibitem{art:97LPO}
M. Laso, M. Picasso, H. C. \"Ottinger,
``2-D time-dependent viscoelastic flow calculations using CONNFFESSIT'',
AIChE J. {\bf 43}, 877 (1997).
%
\bibitem{art:99DEO}
M. Dressler, B. J. Edwards, \"Ottinger,
``Macroscopic thermodynamics of flowing polymeric liquids'',
Rheol. Acta {\bf 38}, 117 (1999).
%
\bibitem{art:03EE}
W. E and B. Engquist,
``The heterogeneous multi-scale methods'',
Comm. Math. Sci. {\bf 1}, 87 (2003).
%
\bibitem{art:05RE}
W. Ren and W. E,
``Heterogeneous multiscale method for the modeling of complex fluids and micro-fluidics'',
J. Compt. Phys. {\bf 204}, 1 (2005).
%
\bibitem{art:11MD}
M. \"Muller and K. C. Daoulas,
``Speeding Up Intrinsically Slow Collective Processes in Particle Simulations by Concurrent Coupling to a Continuum Description'',
Phys. Rev. Lett. {\bf 107}, 227801 (2011).
%
\bibitem{art:13BLR}
M. K. Borg, D. A. Lockerby, J. M. Reese,
``A multiscale method for micro/nano flows of high aspect ratio'',
J. Compt. Phys. {bf 233}, 400 (2013).
%
%
\bibitem{art:03KGHKRT}
I. G. Kevrekidis, C. W. Gear, J. M. Hyman, P. G. Kevrekidis, O. Runborg, and C. Theodoropoulos,
``Equation-free, coarse-grained multiscale computation: enabling microscopic simulations to perform system-level analysis'',
Comm. Math. Sci. {\bf 1}, 715 (2003).
%
\bibitem{art:09KS}
I. G. Kevrekidis and G. Samaey,
``Equation-free multiscale computation: algorithms and applications'',
Annu. Rev. Phys. Chem. {\bf 60}, 321 (2009).
%
\bibitem{art:06DFSKK}
S. De, J. Fish, M. S. Shephard, P. Keblinski, and S. K. Kumar,
``Multiscale modeling of polymer rheology'',
Phys. Rev. E {\bf 74}, 030801(R) (2006).
%
\bibitem{art:13D}
S. De,
``Computational study of the propagation of the longitudinal velocity in a polymer melt contained within a cylinder using a scale-bridging method'',
Phys. Rev. E {\bf 88}, 052311 (2013).
%
%
%
\bibitem{MT2010}
T. Murashima and T. Taniguchi,
``Multiscale Lagrangian Fluid Dynamics Simulation for Polymeric Fluid''
J. Polym. Sci. B {\bf 48}, 886 (2010).
%
\bibitem{MT2011}
T. Murashima and T. Taniguchi,
``Multiscale Simulation of History Dependent Flow in Polymer Melt'',
Europhys. Lett. {\bf 96}, 18002 (2011).
%
\bibitem{MT2012}
T. Murashima and T. Taniguchi,
``Multiscale Simulation of History Dependent Flow in Polymer Melt'',
J. Phys. Soc. Jpn. {\bf 81}, SA013 (2012).
%
\bibitem{art:08YY}
S. Yasuda and R. Yamamoto,
``A model for hybrid simulation of molecular dynamics and computational fluid dynamics'',
Phys. Fluids {\bf 20}, 113101 (2008).
%
\bibitem{art:09YY}
S. Yasuda and R. Yamamoto,
``Rheological properties of polymer melt between rapidly oscillating plates: an application of multiscale modeling'',
Europhys. Lett. {\bf 86}, 18002 (2009).
%
\bibitem{art:10YY}
S. Yasuda and R. Yamamoto,
``Multiscale modeling and simulation for polymer melt flows between parallel plates'',
Phys. Rev. E {\bf 81}, 036308 (2010).
%
\bibitem{art:11YY}
S. Yasuda and R. Yamamoto,
``Dynamic rheology of a supercooled polymer melt in nonuniform oscillating flows between rapidly oscillating plates'',
Phys. Rev. E {\bf 84}, 031501 (2011).
%
\bibitem{art:13MYTY}
T. Murashima, S. Yasuda, T. Taniguchi, and R. Yamamoto,
``Multiscale modeling for polymeric flow: particle-fluid bridging scale methods'',
J. Phys. Soc. Jpn. {\bf 82}, 012001 (2013).
%
\bibitem{art:14YY}
S. Yasuda and R. Yamamoto,
``Synchronized molecular-dynamics simulation via macroscopic heat and momentum transfer'',
Phys. Rev. X {\bf 4}, 041011 (2014).
%
\bibitem{art:90KG}
K. Kremer and G. S. Grest,
``Dynamics of entabgled linear polymer melts: A molecular-dynamics simulation,''
J. Chem. Phys. {\bf 92}, 5057 (1990).
%
\bibitem{art:90BB}
B.H.A.A. van den Brule and S.B.G. O'Brien,
``Anisotropic conduction of heat in a flowing polymeric material'',
Rheol Acta {\bf 29}, 580 (1990).
%
\bibitem{art:96OP}
H. C. \"Ottinger and F. Petrillo,
``Kinetic theory and transport phenomena for a dumbbell model under nonisothermal
conditions'',
J. Rheol. {\bf 40}, 857 (1996).
%
\bibitem{art:96BC}
R. B. Bird and C. F. Curtiss,
``Nonisothermal polymeric fluids'',
Rheol. Acta {\bf 35}, 103 (1996).
%
\bibitem{art:97BCB}
R. B. Bird, C. F. Curtiss, and K. J. Beers,
``Polymer contribution to the thermal conductivity
and viscosity in a dilute solution'',
Rheol. Acta {\bf 36}, 269 (1997).
%
\bibitem{art:01VSIGB}
D. C. Venerus, J. D. Schieber, H. Iddir, J. Guzm\'an, and A. Broerman,
``Anisotropic Thermal Diffusivity Measurements in
Deforming Polymers and the Stress-Thermal Rule'',
Int. J. Thermophysics {\bf 22}, 1215 (2001).
%
\bibitem{art:04SVBBS}
J. D. Schieber, D. C. Venerus, K. Bush, V. Balasubramanian, and S. Smoukov,
``Measurement of anisotropic energy transport in flowing polymers by using a holographic technique'',
PNAS {\bf 101}, 13142 (2004).
%
\bibitem{art:12SVG}
J. D. Schieber, D. C. Venerus, and S. Gupta,
``Molecular origins of anisotropy in the thermal conductivity of deformed polymer melts: stress versus orientation contributions''.
Soft Matter {\bf 8}, 11781 (2012).
%
\bibitem{art:13GSV}
S. Gupta, J. D. Schieber, and D. C. Venerus,
``Anisotropic thermal conduction in polymer melts in uniaxial elongation flows'',
J. Rheol. {\bf 57}, 427 (2013).
%
\bibitem{art:08PMM}
C. J. Pipe, T. S. Majmudar, and G. H. McKinley,
``High shear rate viscometry'',
Rheol Acta {\bf 47}, 621 (2008).
%
\bibitem{art:02YO}
R. Yamamoto and A. Onuki,
``Dynamics and rheology of a supercooled polymer melt in shear flow,''
J. Chem. Phys. {\bf 117}, 2359 (2002).

\end{thebibliography}


\end{document}